\documentclass{Interspeech2024}

\interspeechcameraready

\title{ Signal processing algorithm effective for sound quality of \\ hearing loss simulators  }

\name[affiliation={1}]{Toshio}{Irino}
\name[affiliation={1}]{Shintaro}{Doan}
\name[affiliation={1}]{Minami}{Ishikawa}

\address{
  $^1$Faculty / Graduate School of Systems Engineering, Wakayama University, Japan}
\email{irino@wakayama-u.ac.jp, doan.shintaro@g.wakayama-u.jp, ishikawa.minami@g.wakayama-u.jp}


\usepackage{subcaption}

\begin{document}

\maketitle

\begin{abstract}

Hearing loss (HL) simulators, which allow normal hearing (NH) listeners to experience HL, have been used in speech intelligibility experiments, but not in sound quality experiments due to perceptible distortion. If they produced less distortion, they might be useful for NH listeners to evaluate the sound quality of, for example, hearing aids. We conducted perceptual sound quality experiments to compare the Cambridge version of HL simulator (CamHLS) and the Wakayama version of the HL simulator (WHIS), which has the two algorithms of filterbank analysis synthesis (FBAS) and direct time-varying filter (DTVF). The experimental results showed that WHIS with DTVF produces less perceptible distortion in speech sounds than CamHLS and WHIS with FBAS, even when the nonlinear process is working. This advantage is mainly due to the use of the DTVF algorithm, which could be applied to various signal synthesis applications with filterbank analysis.

\end{abstract}

\section{Introduction}
\vspace{-3pt}
In order to develop the next generation of assistive listening devices that can compensate for the difficulties experienced by elderly people with hearing impairment (HI), it is essential to clarify the dysfunctions. Many psychoacoustic experiments have been conducted by using relatively simple stimulus sounds, such as sinusoids and noise\cite{moore2013introduction}.  In addition, many speech sound experiments have been performed, although they have mainly been limited to intelligibility tests, such as the speech-in-noise test. 
However, it is not easy to determine the degradation factors, whether they are located in the periphery, within the auditory pathway, or due to cognitive factors, because of the considerable variability among HI listeners.
To partially resolve this problem,
a hearing loss (HL) simulator was developed to specify the effects of peripheral dysfunction, such as  absolute threshold (AT) elevation and loudness recruitment \cite{moore2013introduction}, on speech intelligibility \cite{villchur1974simulation}.
Normal-hearing (NH) listeners could evaluate the speech intelligibility of the HL-simulated sounds, which may be similar to what HI listeners perceive.

The development of such HL simulators has a long history.
 The most popular was developed by Moore's group at Cambridge University~\cite{moore1993simulation, nejime1997simulation} (hereafter CamHLS), which has been used in many psychoacoustic studies~\cite{baer1993effects,baer1994effects, glasberg1986auditory,stone1999tolerable}.
There are many other HL simulators, e.g.~\cite{zurek2007hearing, hu2011simulation, grimault2018real,cuevas20183d, mourgela2020investigation}, 
which are not commonly used in psychoacoustic experiments. 
With the goal of being used in various psychoacoustic experiments, an HL simulator was also developed by our group at Wakayama University~\cite{irino2013accurate, nagae2014hearing, irino2023hearing} (hereafter WHIS).
WHIS has been used for several speech perception experiments \cite{matsui2016effect,irino2022speech}.
Deutch and Fels \cite{deutsch2023evaluating} evaluated several HL simulators~\cite{grimault2018real,cuevas20183d,mourgela2020investigation} and showed that WHIS simulated the audiogram better.

The psychoacoustic experiments using the HL simulators were mainly limited to speech intelligibility tasks due to the distortion caused by the nonlinear signal processing used to simulate the HL.  Although the distortion is unavoidable in nonlinear signal processing, it is desirable to reduce the audible distortion to expand the range of experiments, such as speech and sound quality tasks. For example, the sound quality of hearing aids could be tested by NH listeners using the HL simulator with less distortion. 
However, it seems that the sound quality of HL simulator has not yet been evaluated, at least not psychoacoustically. Instead, the spectral distortion between the auditory model outputs of the HI condition and of the NH condition with the HL simulator was evaluated, and the latest version of WHIS~\cite{irino2023hearing} was shown to outperform CamHLS. 
However, it is still uncertain whether the spectral distortion is directory corresponding to the sound quality.

In this paper, we evaluate the speech and instrument sound quality of CamHLS and WHIS through subjective listening tests. They employ different signal processing algorithms. Based on the perceptual results, we discuss signal processing algorithm effective for sound quality of HL simulators.


\vspace{-6pt}
\section{Algorithms of HL simulators}
\label{sec:AlgorithmHLS}
\vspace{-3pt}   

This section briefly summarizes the signal processing of the CamHLS and WHIS to clarify the algorithm differences.


\begin{figure}[t]
  \vspace{-30pt}
  \centering
  \includegraphics[width=1.15\linewidth, bb=0 0 894 942]{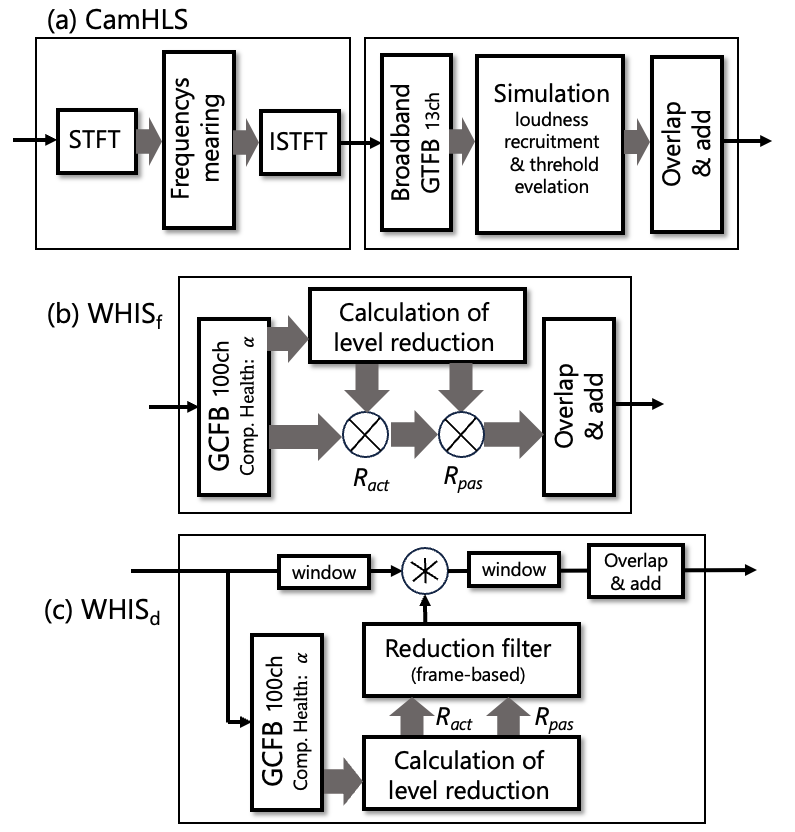}
  \vspace{-12pt}
   \caption{   \label{fig:BlockDiagramHLS}
 Block diagrams of hearing loss simulators. See Section~\ref{sec:AlgorithmHLS} for detail.}
  \vspace{-15pt}
\end{figure}

\begin{figure}[t]
  \centering
  \includegraphics[width=0.9\linewidth, bb=0 0 644 504]{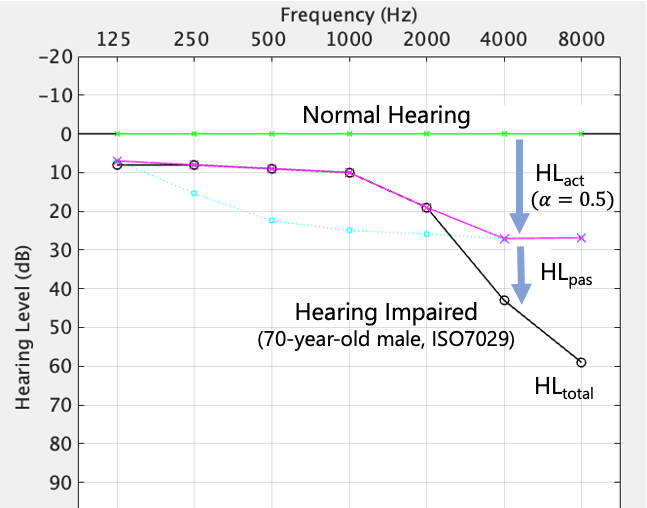}
  \vspace{-2pt}
   \caption{ \label{fig:AudiogramHL} Relationship between the hearing level, the active HL, and the passive HL in the audiogram (adapted from the GUI version of WHIS~\cite{irino2023hearing}). The green line shows the hearing level of NH; the black line shows the average hearing level of the 70-year-old male  (70-yr) listeners~\cite{iso7029}; the magenta line represents the active HL, $HL_{act}$, that appears on the audiogram.  The difference between the magenta and black lines is the passive HL, $HL_{pas}$.}
  \vspace{-5pt}
\end{figure}

\vspace{-6pt}
\subsection{CamHLS}
\label{sec:AlgorithmCamHLS}
\vspace{-6pt}

Figure ~\ref{fig:BlockDiagramHLS}(a) shows the block diagram of CamHLS. There are two stages: the frequency smearing stage using the Short-Time Fourier Transform (STFT) and the stage for simulating loudness recruitment and absolute threshold elevation. These are basically an analysis/synthesis method based on frequency component decomposition and summation with overlap-and-add. The frequency smearing process in the first stage is introduced to simulate bandwidth widening in HI listeners~\cite{baer1993effects,baer1994effects}. This process produces perceptual distortion (i.e.,`grittiness' sound as described in the CamHLS software). This makes difficult to use CamHLS in sound quality experiments.
The algorithm in the second stage is based on a loudness model ~\cite{moore1997model} proposed to reflect the HL caused by the dysfunction of outer hair cell (OHC, $HL_{OHC}$) and inner hair cell (IHC, $HL_{IHC}$). They defined that the total HL, $HL_{total}$, that appears on the audiogram is assumed to be their sum on a dB scale:
\setlength{\abovedisplayskip}{2pt} 
\setlength{\belowdisplayskip}{2pt} 
 \begin{equation}
   HL_{total} = HL_{OHC}+HL_{IHC}.
    \label{eq:HLtotal_OHC+IHC}
\end{equation}
Since the HL may not be caused by OHC and IHC alone, we have preferred to use active HL ($HL_{act} \ge 0 $) and passive HL ($HL_{pas} \ge 0 $) as 
 \begin{equation}
   HL_{total} = HL_{act}+HL_{pas}.
    \label{eq:HLtotal_ACT+PAS}
\end{equation}


\subsection{WHIS}
\label{sec:AlgorithmWHIS}
Figure~\ref{fig:BlockDiagramHLS}(b) shows WHIS using the FilterBank Analysis/Synthesis (FBAS) method (hereafter, $\rm WHIS_f$) as in CamHLS. 
Figure~\ref{fig:BlockDiagramHLS}(c) shows WHIS using the Direct Time-Varying Filter (DTVF) method (hereafter, $\rm WHIS_d$) ~\cite{irino2022speech, irino2023hearing}. 

\vspace{-6pt}
\subsubsection{Analysis algorithm}
\vspace{-6pt}
The analysis part is the same for both types.
The gammachirp filterbank (GCFB) is used to reflect the dysfunction of the active cochlear process via a parameter called compression health, $\alpha$. The HL simulation is performed by calculating the level reduction of the active process, $R_{act}$, and the passive process, $R_{pas}$, in the cochlear input-output function. $R_{act}$ depends on the input sound level and can be written as $R_{act}(f_c, L(\tau))$, where $L(\tau)$ is the estimated level at frame time $\tau$ in a filter frequency, $f_c$. $R_pas(f_c)$ is a constant. The total level reduction $R_{total}$ is represented as the sum of these values on a dB scale: 
 \begin{equation}
     R_{total}(f_c, L(\tau)) = R_{act}(f_c, L(\tau))+R_{pas}(f_c).
    \label{eq:Rtotal_Ract+Rpas}
\end{equation}
It is similar to Eq.~\ref{eq:HLtotal_ACT+PAS}. In fact, at a  filter frequency $f_c$, $R_{pas}=HL_{pas}$ because it is a constant and $R_{act}(L_{AT}) = HL_{act}$ where $L_{AT}$ is the absolute threshold (AT) level.

Figure~\ref{fig:AudiogramHL} illustrates the relationship between $HL_{total}$, $HL_{act}$, and $HL_{pas}$ on the audiogram. In this case, $HL_{total}$ is the average hearing level of 70-year-old male (hereafter, 70-yr) listeners~\cite{iso7029}. $HL_{act}$ ($\alpha = 0.5$) increases the hearing level up to the $HL_{total}$ line. $HL_{pas}$ fills the gap between $HL_{total}$ and $HL_{act}$. If $\alpha = 1$, the cochlear active process is healthy and $HL_{act}=0$. Therefore, $HL_{total}=HL_{pas}$ and it becomes a constant level reduction that can be simulated by a linear filter. If $\alpha < 1$, an nonlinear level-dependent filtering is required. When $\alpha = 0$, the active process is completely damaged, $HL_{act}$ becomes maximum and $HL_{pas}$ becomes relatively smaller. 

Table\ref{tab:HL} shows how the 70-yr hearing level is divided into the sum of $HL_{act}$ and $HL_{pas}$ depending on the value of $\alpha$. 
Note that the upper limit of $HL_{act}$ is $HL_{total}$ even if $\alpha$ is set to a low value in the software. For example, $HL_{act}$+$HL_{pas}$ = 8+0 at 125~Hz with $\alpha=0.5$ and $\alpha=0$. The difference between the $\alpha$ values of 0.5 and 0 appears in this case at high frequencies (4000 and 8000~/Hz).


\begin{table}[t]
 \tabcolsep = 4pt 
 \caption{Average hearing level (dB) of 70-yr~\cite{iso7029} (i.e.,  $HL_{total}$). The lower three rows show the sum of the active and passive level reductions by WHIS on the audiogram: $HL_{act}$ + $HL_{pas}$, which are dependent on the $\alpha$ values.}
 \vspace{-5pt}

 \label{tab:HL} 
 \centering

  \begin{tabular}{|c||c|c|c|c|c|c|c|c|c|}
   \hline
    Freq. & 125 & 250 & 500 & 1000 & 2000 & 4000 &  8000\\
   \hline
    70-yr & 8 & 8 & 9 & 10 & 19  & 43  & 59 \\
     \hline 
$\alpha=1$& 0+8 & 0+8 & 0+9 & 0+10 & 0+19  & 0+43  & 0+59 \\    
$\alpha=0.5$& 8+0 & 8+0 & 9+0 & 10+0 & 19+0  & 27+16  & 27+32 \\
$\alpha=0$ & 8+0 & 8+0 & 9+0 & 10+0 & 19+0  & 43+0   & 44+15 \\
     \hline  
  \end{tabular}
    \vspace{-15pt}
\end{table}


\vspace{-6pt}
\subsubsection{Synthesis algorithm}
\label{sec:Synthesis}
\vspace{-6pt}
The synthesis procedures of $\rm WHIS_f$ (Fig.~\ref{fig:BlockDiagramHLS}(b)) and $\rm WHIS_d$ (Fig.~\ref{fig:BlockDiagramHLS}(c)) are completely different. $\rm WHIS_f$ uses a common filterbank analysis/synthesis method. It is simple and widely used. However, it is important to align the phase lag, which depends on filter characteristics such as frequency and bandwidth, before performing the overlap and add (e.g.~\cite{irino1999analysis}). Incomplete alignment can cause distortion.
$\rm WHIS_d$ was proposed to avoid this problem.
A minimum phase filter which corresponds the frequency distribution of $R_{total}(f_c)$ is calculated for each time frame $\tau$.  The filter is convolved with the windowed input signal and then overlapped and added frame by frame.
This method seems to be advantageous for sound quality because there is a single filter applied for each frame and it is not necessary to compensate the phase lag of multiple filters.
However, it has not been confirmed in perceptual sound quality experiments.


\vspace{-3pt}
\section{Perceptual experiments}
\label{sec:PerceptualExp}
\vspace{-3pt}
Subjective listening tests were conducted using a paired comparison test that is well established in psychoacoustics because our goal was to evaluate HL simulators that can be used in formal experiments.
MUSHRA~\cite{MUSHRA} is not suitable for this purpose because it was designed to quickly assess the quality of sound encoding, but sacrifices accuracy and reliability, as stated in the recommendation itself. 
In addition, MUSHRA requires a hidden reference of the original or ``high quality'' sample and a hidden anchor filtered by a 3.5kHz low-pass filter. These requirements are unreasonable for the current evaluation because the sound pressure levels of the original samples are different and the HL simulation already has a low-pass characteristic.


\vspace{-3pt}
\subsection{Sound material}
\label{sec:SpeechMatrial}
\vspace{-3pt}

\subsubsection{Speech sounds}
\label{sec:SourceSpeech}
\vspace{-3pt}
We used the male voice from the FW07 database for Japanese four-mora (roughly four-syllable) words with word familiarity control~\cite{sakamoto2006new,kondo2011spoken}. The words with the highest familiarity were used for this experiment, as no lexical judgement was made. We randomly selected 32 words: 10 for the main experiment, 10 for the practice session, and 12 for the training session. After preliminary listening, we decided to apply a room impulse response to the word sounds to make it easier to detect the distortion. Moreover, people with HI usually hear sounds in everyday environments, not just dry sources. The impulse response was derived from the database of the University of Aachen ~\cite{jeub2009binaural}. The conditions were an office room, speaker to microphone distance of 3m and a reverberation time of 0.48s. The duration of the word sounds was about 0.8~s.


\subsubsection{Instrument sounds}
\label{sec:SourceInstrument}
\vspace{-3pt}
The original music consisted of simple three consecutive triads (three-note chords) with the same instrument to allow the listener to focus on detecting the distortion.  The duration of each chord was 1~s and the total stimulus was 3~s. Four instrument sounds were created in GarageBand (Apple): Tuba, Grand Piano, Cheap Organ, and French Horn. The three triads in Roman numeral analysis are $I^6-V^{+6}-I^6$ chords in the major key ~\cite{kostka@tonal}. The highest note of these chords was F3 for Tuba and  C4 for the other instruments. These four stimuli were used for the main session. In addition, tones with one or more lower semitones were created for the training and practice sessions.
Note that the total duration of the triads was 3~s, which was longer than that of the speech sounds, so the number of sound pairs was reduced to keep the duration of the experiment within a reasonable range.

\begin{figure*}[t]
\begin{tabular}{cc}
   \begin{minipage} {0.5\hsize}
  \centering
  \includegraphics[width=\linewidth]{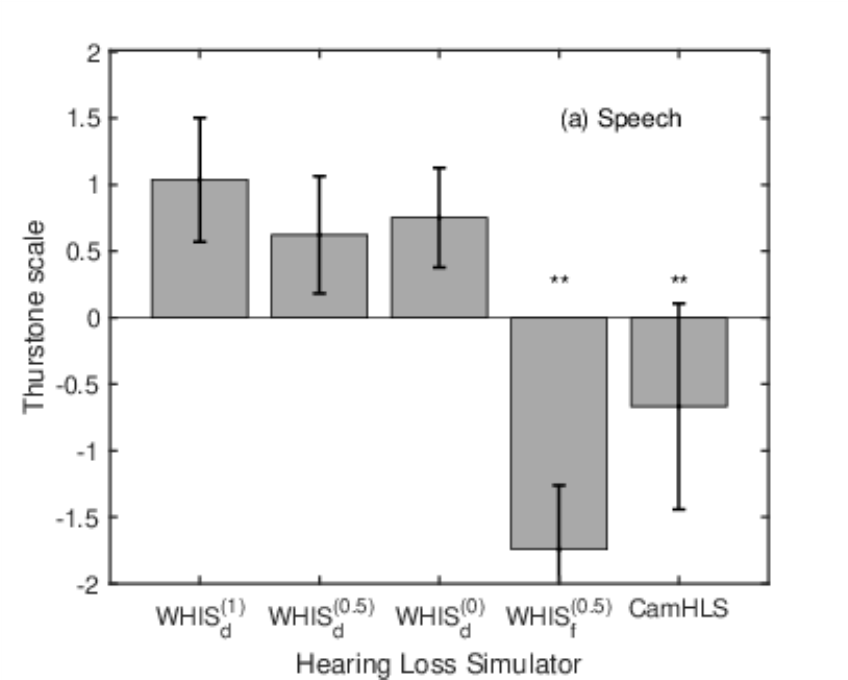}
  \label{fig:bar_speech}
  \end{minipage} 

   \begin{minipage} {0.5\hsize}
  \includegraphics[width=\linewidth]{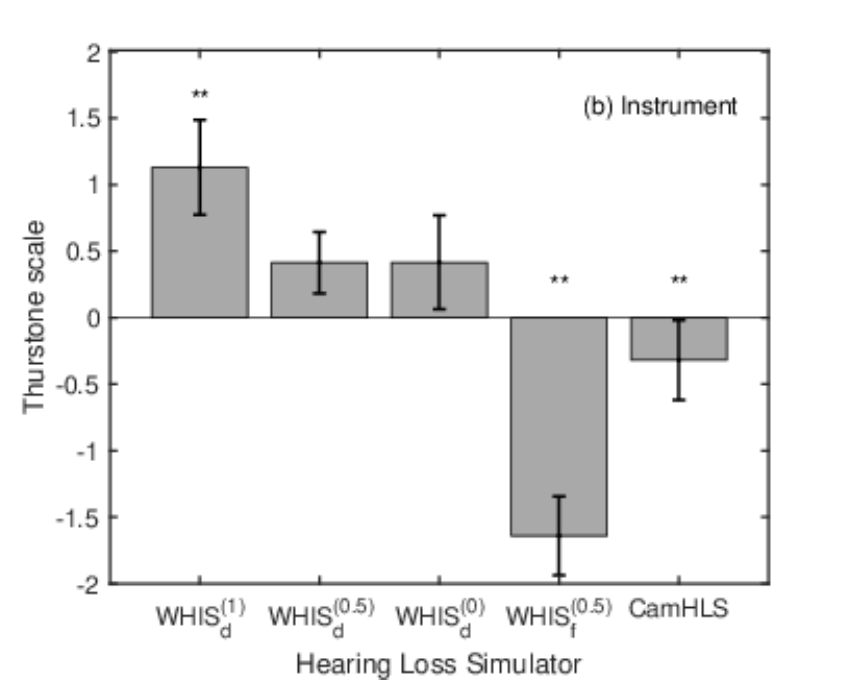}
      \label{fig:bar_instrument}
     \end{minipage} 
\end{tabular}
\vspace{-15pt}
   \caption{\label{fig:Result_BarError} Mean (bar) and 99\% confidence interval (error bar).  Double asterisks (**) indicate the HL conditions whose means were significantly different from $\rm WHIS_d^{(0.5)}$ using Tukey's HSD tests at the 0.01 significance level.}
\vspace{-12pt}
\end{figure*}


\subsubsection{HL simulation}
\label{sec:HearingLossSimulation}
\vspace{-3pt}
The speech and instrument sounds were processed using CamHLS, $\rm WHIS_d$, and $\rm WHIS_f$ with the 70-yr condition shown in Table~\ref{tab:HL}.  
The reason for choosing this condition was to ensure effective experiments by balancing the increase in nonlinear distortions and the decrease in the audibility of small distortions with simulated age. Although the condition is limited due to the time-consuming paired comparison experiments, the results would be valid for other HL profiles since the theoretical study has already shown that the spectral distortion measures for a wide range of HL profiles are smaller for WHIS than for CamHLS~\cite{irino2023hearing}. 
We have used the CamHLS default settings as supplied with the software. For $\rm WHIS_d$ and $\rm WHIS_f$, it is possible to control the degree of dysfunction of the nonlinear active process by the $\alpha$ value, as described in Section~\ref{sec:AlgorithmWHIS}.
After some preliminary listening, we set  ${\rm WHIS_d}$ to $\alpha$ values of 1, 0.5, and 0  (hereafter, $ \rm WHIS_d^{(1)},WHIS_d^{(0.5)},WHIS_d^{(0)}$) and ${\rm WHIS_f}$ to the $\alpha$ value of 0.5 ({$\rm WHIS_f^{(0.5)}$}). 
Note that $\rm WHIS_d^{(1)}$ is virtually equivalent to a linear low-pass filter with the 70-yr hearing level frequency response as described in Section~\ref{sec:AlgorithmWHIS}. Therefore, there is no nonlinear distortion due to the HL simulation.
As the $\alpha$ value decreases, the distortion increases due to the effect of the nonlinear process. Our concern was how much of the nonlinear distortion would be perceived by the listener.

The source sound level was set to $L_{eq}$ of 70~dB to produce sufficient level for subjective judgments of distortion after the HL simulation.
The output levels were dependent on the simulation conditions and could affect the judgments. Therefore, the sound levels for each word and instrument were normalized to the level processed by $\rm WHIS_d^{(1)}$.
The final levels were approximately the same: $L_{eq}$ was 60.8 (SD $\pm 0.1$) dB for speech and 61.1 (SD $\pm 0.2$) dB for instrument. 


\subsection{Experimental procedure}
\label{sec:SbjSI_Procedure}
\vspace{-3pt}

The sound quality evaluation was performed using a paired comparison method. 
The speech experiment was conducted first, followed by the instrument experiment for the same participants.
The presentation of the sound pairs and the response collection were performed using a set of web pages ~\cite{yamamoto2021comparison}. 
Participants had to respond to the interval in which the sound contained more distortion. 
For each of the speech and instrument sounds, all combinations and their reverse order of pair sounds for five HL simulation conditions ($_5C_2 \times 2 = 20$) were prepared. Thus, in the main experiment, the participants judged 200 pairs of 10 word sounds and 80 pairs of 4 instruments in random order. 

\subsubsection{Training and practice sessions}
\label{sec:Training_Practice}
\vspace{-3pt}

It was necessary to familiarize the participants with the sound quality judgment because they were naive to such experiments.
In the training session for the speech experiment, we used a total of 12 pairs: 3 pairs of ${\rm WHIS_d^{(1)}} $ and ${\rm WHIS_f}$, 3 pairs of ${\rm WHIS_d^{(1)}} $ and CamHLS, and their reverse order. This is because we clearly perceived nonlinear distortions in ${\rm WHIS_f}$ and CamHLS, while ${\rm WHIS_d^{(1)}}$ contains no nonlinear distortions. In the training session for the instrument experiment, we used a total of 8 pairs due to the relatively long sounds: 4 pairs of ${\rm WHIS_d^{(1)}} $ and ${\rm WHIS_f}$, 4 pairs of ${\rm WHIS_d^{(1)}} $ and CamHLS. Participants judged which sound had more distortion and received feedback from the collect response. 
After listening to the all the sound pairs, the test score was calculated and the participants who scored 83\% (=10/12) for speech and 88\% (=7/8) for instrument passed this session and proceeded to the practice session.  If they did not pass, the training session was repeated.
In the practice session, they listened to sound pairs that were the same as in the main session, except that different words or different sounds of the same instruments were used. There were 20 speech sound pairs and 10 instrument sound pairs.



\subsubsection{Participants and audio settings}
\label{sec:Participants_Audio}
\vspace{-3pt}
The experiment was conducted with fifteen young NH listeners (aged 21--23 years) who read information about the experiments and gave informed consent. The experiments were approved by the Ethics Committee of Wakayama University (No. 2015-3, Rei01-01-4J, and Rei02-02-1J).
Participants were seated in a sound-attenuated room with a background noise level of approximately 26dB in $L_{\rm Aeq}$. They had a hearing level of less than 20~dB between 125~Hz and 8,000~Hz. 
The sounds were presented diotically through a DA-converter (SONY, NW-A55) via headphones (SONY, MDR-1AM2). 
The sound was presented as 48~kHz and 16-bit wav files through the web page using Google Chrome.
Level calibration was performed using an artificial ear (Br\"{u}el \& Kj\ae r, Type~4153), a microphone (Br\"{u}el \& Kj\ae r, Type~4192),  and a sound level meter (Br\"{u}el \& Kj\ae r, Type~2250-L).


\color{black}

\subsection{Result}
\label{sec:ResultPerceptualExp}
\vspace{-3pt}

For each listener and for each HL simulator condition, responses were aggregated for 10 words and for 4 instruments.
Scores on the Thurstone scales were calculated from these. 
Figure~\ref{fig:Result_BarError} shows the mean and 99\% confidence interval across listeners for the speech (a) and instrument (b) experiments. The perceived distortion is smaller when the score is higher. The trend across the HL simulator conditions was similar in both experiments. Mean scores were positive in the three $\rm WHIS_d$ conditions and negative in $\rm WHIS_f$ and CamHLS.
The score for $\rm WHIS_f^{(0.5)}$ is the most negative, meaning that the distortion was perceived the most.


Tukey's HSD tests were performed at a significance level of 0.01. Double asterisks (**) in Fig.~\ref{fig:Result_BarError}(a)(b) indicate the HL conditions whose means were significantly different from $\rm WHIS_d^{(0.5)}$. There were significant differences between $\rm WHIS_d^{(0.5)}$ and $\rm WHIS_f^{(0.5)}$ and between $\rm WHIS_d^{(0.5)}$ and CamHLS in the both experiments. This was also the case for $\rm WHIS_d^{(1)}$ and $\rm WHIS_d^{(0)}$. 
In the speech experiment (Fig.~\ref{fig:Result_BarError}(a)), there were no significant difference between $\rm WHIS_d^{(1)}$, $\rm WHIS_d^{(0.5)}$, and $\rm WHIS_d^{(0)}$. This means that the distortion was not very noticeable even as the $\alpha$ value decreased and the nonlinear process was in operation.
In contrast, in the instrument experiment (Fig.~\ref{fig:Result_BarError}(b)), there were significant differences between $\rm WHIS_d^{(1)}$ and  $\rm WHIS_d^{(0.5)}$, as indicated by **, and also between $\rm WHIS_d^{(1)}$ and $\rm WHIS_d^{(0)}$. 
The nonlinear distortion became noticeable when the value of $\alpha$ was less than 1, i.e. with the nonlinear process.  
However, these distortions were significantly less pronounced than those of $\rm WHIS_d^{(0.5)}$ and CamHLS. 
In summary, WHISd was found to be generally superior to WHISf and CamHLS.




\vspace{-3pt}
\section{Effective algorithm for sound quality}
\label{sec:AlgorithmQuality}
\vspace{-3pt}
Here we discuss the relationship between the algorithms of the HL simulators and the results of the sound quality assessment.
Figure~\ref{fig:BlockDiagramHLS} in Section~\ref{sec:AlgorithmHLS} show the algorithms of CamHLS (a), $\rm WHIS_d$ (b), and $\rm WHIS_f$ (c).

As described in Section~\ref{sec:AlgorithmCamHLS}, the degradation of sound quality in CamHLS is mainly caused by the frequency smearing process to simulate the band bandwidth widening in HI listeners. By removing this process, the quality can be improved but the effect of bandwidth widening cannot be simulated anymore.

$\rm WHIS_d$ and $\rm WHIS_f$ use the same analysis algorithm with GCFB and differ only in the synthesis algorithm, as described in Section~\ref{sec:AlgorithmWHIS}.
GCFB can also reflect the effect of bandwidth widening. Although the degree is generally smaller than in CamHLS, the difference is not expected to be as large for the 70-yr simulation.
The sound quality of $\rm WHIS_d$ using DTVF was significantly better than that of $\rm WHIS_f$ using FBAS. 
The degradation factor in FBAS seems to be the misalignment of the phase lag before the overlap-and-add.  
The current implementation compensates for a fixed amount of lag for each filter, although the filter response depends on the $\alpha$ value. Introducing the $\alpha$ dependency can solve the problem, but makes the algorithm more complex.
In contrast, the DTVF algorithm simply applies a single filter per frame, which is calculated from the amplitude frequency response of $R_{total}$, in Eq.~\ref{eq:Rtotal_Ract+Rpas}, which does not require precise phase alignment.
As a result, the DTVF algorithm is advantageous for the sound quality of the HL simulator.


\vspace{-3pt}
\section{Summary}
\label{sec:Summary}
\vspace{-3pt}

We conducted perceptual experiments to evaluate the sound quality of the HL simulators and argued the relationship to the signal processing algorithms.
We compared the widely used CamHLS and the recently proposed $\rm WHIS$, which has two algorithms using DTVF ($\rm WHIS_d$) and FBAS ($\rm WHIS_f$).
The perceived sound quality of $\rm WHIS_d$ was better than that of $\rm WHIS_f$ and CamHLS. $\rm WHIS_d$ seems to produce little perceptible distortion in speech sounds, even when the nonlinear process is working. 
This could be useful for evaluating the speech quality of hearing aids that have more nonlinear distortion.
This advantage is mainly due to the use of the DTVF algorithm, which can be applied to various signal synthesis applications with filterbank analysis, not only to HL simulators.

\vspace{-3pt}
\section{Acknowledgements}
\vspace{-3pt}

\ifinterspeechfinal
    This research was supported by JSPS KAKENHI: Grant Numbers JP21H03468, JP21K19794, and JP24K02961. The authors would like to thank Prof. Toshie Matsui for her help in describing the instrument sounds.
\else
    This research was supported by AAA: Grant Numbers XXX and YYY. The authors would like to thank ZZZ for the help in describing the instrument sounds.
\fi


\bibliographystyle{IEEEtran}
\bibliography{Ref_WHIS_Mar24}

\end{document}